\documentclass[aps,pre,reprint,floatfix]{revtex4-1}

\usepackage{amsmath}
\usepackage{amssymb}
\usepackage{amsfonts}
\usepackage{graphicx}
\usepackage{dcolumn}
\usepackage{bm}
\usepackage{hyperref}
\usepackage{mathptmx}

\begin{document}

\title{Logarithmic finite-size scaling correction to the leading Fisher zeros in the $p$-state clock model:\\A higher-order tensor renormalization group study}

\author{Seongpyo Hong}
\author{Dong-Hee Kim}
\email{dongheekim@gist.ac.kr}
\affiliation{Department of Physics and Photon Science, School of Physics and Chemistry, Gwangju Institute of Science and Technology, Gwangju 61005, Korea
}

\begin{abstract}
We investigate the finite-size-scaling (FSS) behavior of the leading Fisher zero of the partition function in the complex temperature plane in the $p$-state clock models of $p=5$ and $6$. We derive the logarithmic finite-size corrections to the scaling of the leading zeros which we numerically verify by performing the higher-order tensor renormalization group (HOTRG) calculations in the square lattices of a size up to $128 \times 128$ sites. The necessity of the deterministic HOTRG method in the clock models is noted by the extreme vulnerability of the numerical leading zero identification against stochastic noises that are hard to be avoided in the Monte-Carlo approaches. We characterize the system-size dependence of the numerical vulnerability of the zero identification by the type of phase transition, suggesting that the two transitions in the clock models are not of an ordinary first- or second-order type. In the direct FSS analysis of the leading zeros in the clock models, we find that their FSS behaviors show excellent agreement with our predictions of the logarithmic corrections to the Berezinskii-Kosterlitz-Thouless ansatz at both of the high- and low-temperature transitions.
\end{abstract}

\maketitle

\section{Introduction}\label{Intro}

Finite-size-scaling (FSS) analysis is an essential numerical tool 
to study phase transitions and critical phenomena~\cite{Cardy1988}. 
The singular behavior of free energy characterizing a phase transition 
is hidden in finite-size systems available in numerical simulations 
because the growth of the correlation length is governed by 
the finiteness of the system.
For instance, in a system below the upper critical dimension, 
the correlation length is typically considered to be bounded by
the linear dimension of the system~\cite{Fisher1971,Fisher1972} 
while it can exceed the linear dimension in a system 
above the upper critical dimension~\cite{Berche2012,Kenna2013a,Kenna2015}.
Around the transition point, the FSS ansatz relates the scaling behavior 
of an observable to the system size through the limited divergence 
of the correlation length, enabling a precise determination of 
the transition point and critical exponents from the curve collapse 
or the extrapolation with systems of different sizes. 
While the access to larger systems is thus crucial for better FSS analysis, 
the practical limit of an available system size depends on the character 
of the phase transition that the system undergoes as well as 
a particular numerical simulation method to be used.

In this paper, we focus on the FSS behaviors of the leading Fisher zeros 
in the systems undergoing the Berezinskii-Kosterlitz-Thouless (BKT) transitions
\cite{Berezinskii1971,Kosterlitz1972,Kosterlitz1973}.
The zeros of the partition function provide a way to characterize 
phase transitions without defining order parameters 
(for a review, see, for instance, Ref.~\cite{Bena2005}). 
The connection between the singular behavior of the free energy 
and the partition function zeros was formulated 
first in the plane of complex fugacity by Lee and Yang~\cite{Yang1952} 
and then in the plane of complex temperature by Fisher~\cite{Fisher1965}
which we consider here.
While the Fisher zero coincides with the transition point only 
in the thermodynamic limit, the leading zero with the smallest magnitude 
of its imaginary part systematically approaches the real-temperature axis
as the system size increases. 
The FSS behavior of the leading Fisher zero is well established 
in the first- and second-order transitions 
(see, for instance, Ref.~\cite{Janke2001} and references therein).
For the BKT transitions, although the FSS behavior of Lee-Yang 
zeros and the logarithmic corrections was derived and 
numerically examined long ago~\cite{Kenna1995,Irving1996,Kenna1997}, 
progress with Fisher zeros was much slower. 
The Fisher zero study has been extended very recently 
to the BKT transitions in the two-dimensional 
$XY$ model~\cite{Denbleyker2014,Rocha2016,Costa2017}
and the $p$-state clock model~\cite{Hwang2009,Kim2017}.

The Monte Carlo (MC) estimates of the leading Fisher zeros 
have shown limited success in characterizing the BKT transitions. 
In the $p$-state clock model~\cite{Hwang2009,Kim2017}, 
the leading zero calculations based on the Wang-Landau (WL) density of 
states~\cite{Wang2001a,Wang2001b}
turned out to be reliable only up to $L \lesssim 32$ in the square
lattices of $L \times L$ sites. This may be surprising since the system
size reached $L=128$ already a decade ago in the previous Fisher zero study 
on the Potts model~\cite{Alves2002}, implying that the numerical difficulty 
of finding the leading Fisher zero may differ with 
the type of phase transition.
In Ref.~\cite{Kim2017}, it was argued that the nondivergent specific 
heat in the BKT transition was the fundamental origin of 
the small accessible system sizes for the Fisher zero search 
and the consequent inconclusive FSS behavior of the leading zeros. 
In particular, the low-temperature transitions at all $p$'s and 
the high-temperature transition at $p=5$ remain 
uncharacterized with the Fisher zero in the $p$-state clock model.

On the other hand, in the $XY$ model in the square lattices, 
the deterministic calculations by using the higher-order tensor 
renormalization group (HOTRG) method~\cite{Denbleyker2014} 
provided the computation of the leading Fisher zeros 
for the system sizes up to $L=128$. Although the WL approach 
with energy space binning~\cite{Rocha2016,Costa2017} reported 
the leading-zero computation performed for up to $L=200$, 
the transition temperature estimate was $T_\mathrm{BKT} \approx 0.70$,
which deviated from the known value $T_\mathrm{BKT} \approx 0.89$
\cite{Kenna2006,Hasenbusch2005,Komura2012}.
In contrast, the HOTRG calculation~\cite{Denbleyker2014}
for the power-law leading-zero trajectory was consistent 
with the known value of the BKT transition temperature.

The main goal of this paper is to provide a reliable FSS analysis
of the leading Fisher zeros to characterize both of the low- and 
high-temperature transitions in the $p$-state clock model.
While the previous results in the $XY$ model~\cite{Denbleyker2014} 
suggested a numerical advantage of using the HOTRG method, 
we find that it still needs a more precise analytic treatment of 
finite-size effects appearing in the clock model.
We derive the logarithmic corrections to the finite-size scaling of 
the leading zeros, which turns out to be essential to determine 
the transition temperature and understand the particularly strong 
finite-size influences observed at the lower transitions. 
By employing the HOTRG calculations and the logarithmic
correction to the FSS ansatz, we obtain the Fisher-zero estimates
of the transition temperatures that agree well with the previous
estimates obtained from different measures. The same BKT ansatz 
with the logarithmic finite-size correction successfully describes 
the leading zero behaviors at both of the upper and lower transitions
in the $p$-state clock model.

The importance of logarithmic corrections at the BKT transitions was 
pointed out in the seminal studies of Lee-Yang zeros 
in the two-dimensional $XY$ and step models 
\cite{Kenna1995,Irving1996,Kenna1997}.
In the second-order transitions, the scaling relations between 
logarithmic correction exponents were derived through the behaviors 
of Lee-Yang and Fisher zeros~\cite{Kenna2006a,Kenna2006b,Kenna2013}. 
While the behavior of Fisher zeros at the BKT transition was not considered 
in these previous works, our FSS analysis based on the HOTRG calculations 
provides the numerical evidence of the logarithmic scaling behavior 
of the leading Fisher zeros characterizing the two BKT transitions 
in the $p$-state clock model.

In addition, we revisit the numerical advantage of the HOTRG 
calculations over the MC estimates when studying the leading Fisher 
zeros in the systems undergoing the BKT transitions. 
By performing analytic and numerical analysis on the hill-valley 
structure of the partition function around the leading zero location, 
we find that the tolerance to the noises for the visual identification 
of the zero, which we call ``numerical visibility,'' 
shows the distinct FSS behavior that encodes the character
of the associated phase transition. 
In the first-order transition, the numerical visibility of the zero 
location under the finite noises does not decrease with increasing 
system size, making the leading zeros well accessible 
with the MC estimates in a large system.
On the other hand, in the second-order transition, the visibility 
can decay slowly, depending on the criticality of the specific heat. 
In the BKT transition, the leading zero becomes exponentially 
less visible as the system size increases, indicating that 
an extremely accurate computation of the partition function 
is required for the systematic FSS analysis of the leading zeros.

This paper is organized as follows. 
In Sec.~\ref{sec:method}, we describe the numerical procedures 
of finding the Fisher zeros and provide the numerical details 
of the WL and HOTRG methods to evaluate the partition functions at 
complex temperatures. We present our main results in two parts. 
In Sec.~\ref{sec:error}, we derive the system-size dependence 
of the numerical visibility of the leading zero for different 
types of phase transition and demonstrate it
in the Ising, Potts, and clock models.
In Sec.~\ref{sec:clock}, we introduce the logarithmic corrections
to the BKT ansatz to derive the FSS forms of the leading zeros.
We perform the analysis with the HOTRG data in the five- and six-state
clock models to locate the transition points and discuss 
the BKT character of the zeros at the upper and lower transitions.
The summary and conclusions are given in Sec.~\ref{sec:summary}.

\section{Models and Numerical Methods}
\label{sec:method}

\subsection{Models for different phase transitions}

While the Fisher-zero characterization of the $p$-state clock model
is of our main interest, we also consider the other well-known 
classical spin models in the square lattices to compare the numerical 
difficulty of finding the leading Fisher zero between the different 
types of phase transition.
For the ordinary first-order and second-order transitions, we consider 
the spin-1/2 Ising model and the $q$-state Potts model with $q=3$ and $q=10$. 
The Ising Hamiltonian is given as $H = -\sum_{\langle i,j \rangle}s_i s_j$ 
where the spin variable $s_i$ at site $i$ takes the values of $\pm 1$, 
and the summation runs over the nearest neighbor sites. 
The $q$-state Potts model is described by 
the Hamiltonian $H = -\sum_{\langle i,j \rangle}\delta_{\sigma_i,\sigma_j}$ 
where the Potts spin takes $\sigma = 0, \ldots, q-1$. The $10$-state Potts
model is used as an example system undergoing the first-order transition.
The three-state Potts and the Ising models exemplify 
the second-order transitions with different critical exponents.

The Hamiltonian of the clock model is given as 
$H = -\sum_{\langle i,j \rangle} \cos (\theta_i - \theta_j)$, 
where the spin angle $\theta = 2\pi n / p$ has a discrete value 
with an integer $n \in \{ 0, \ldots, p-1 \}$. 
The $p$-state clock model is a cousin of the $XY$ model with discrete 
$\mathrm{Z}(p)$ symmetry. Despite the discrete symmetry, 
it was analytically found that the continuous $\mathrm{U}(1)$ symmetry
would emerge at $p>4$, and the massless intermediate-temperature 
phase would undergo two BKT transitions into the low-temperature 
ordered and high-temperature disordered phases
\cite{book:Jose,Elitzur1979,Cardy1980,Einhorn1980,Hamer1980,Frohlich1981,Nienhuis1984,Ortiz2012}.
The nature of the two transitions in the $p$-state 
clock model has been studied widely with various numerical methods
and different measures~\cite{Tobochnik1982,Challa1986,Yamagata1991,Tomita2002,Borisenko2011,Borisenko2012,Lapilli2006,Baek2010a,Baek2010b,Baek2013,Kumano2013,Chatelain2014,Hwang2009,Kim2017,Chen2017,Chen2018,Surungan2019}.
However, the characteristics of the Fisher zeros remain unclear
particularly for $p=5$ and at the low-temperature transition 
even for a higher $p$.

\begin{figure}[t]
\centering\includegraphics[width=0.48\textwidth]{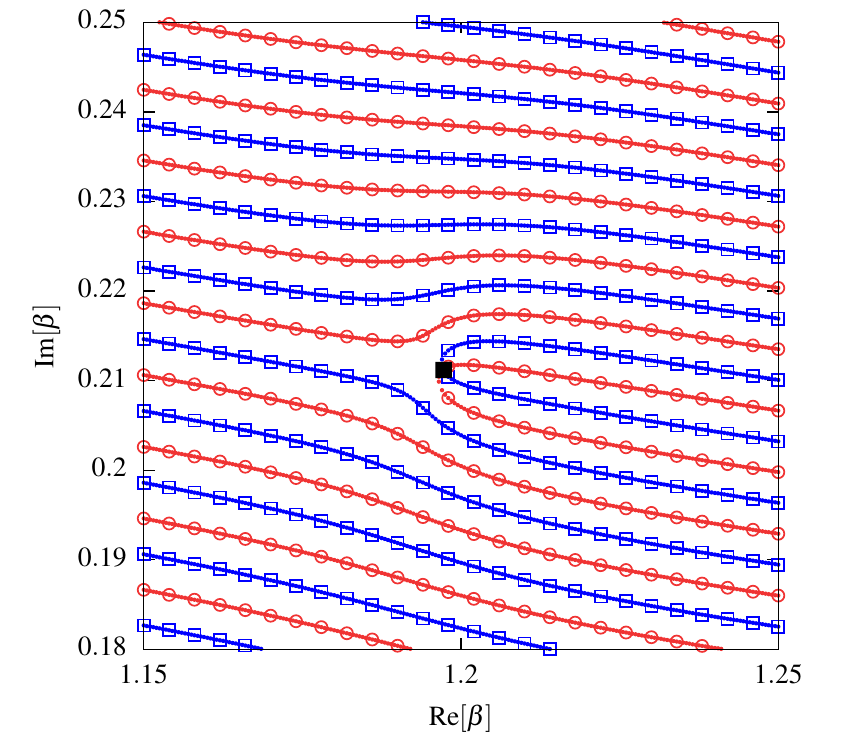}
\caption{\label{fig1}
Map of the zeros of real (square) and imaginary (circle) 
parts of the partition function in the five-state clock model.
The partition function is evaluated with the WL density of 
states sampled in the system of $L=16$.
The Fisher zero is marked with the filled square.
}
\end{figure}

\subsection{Numerical strategies of finding the Fisher zeros}

We follow a standard strategy of searching for
partition function zeros that is usually evaluated with 
the reweighting scheme based on the Monte Carlo 
estimate of spectral densities
\cite{Falcioni1982,Marinari1984,Ferrenberg1988,Ferrenberg1989}.
Although the recipe is well known (for instance, see the procedures in 
Ref.~\cite{Kenna1997} for Lee-Yang zeros and \cite{Alves1992} 
for Fisher zeros and references therein),
let us briefly go through the implementation with the WL density
of states for later discussion on the numerical uncertainty issue 
at the BKT transition given in Sec.~\ref{sec:error}. 
The two-step procedures~\cite{Alves1992} are composed of 
the graphical search to find an approximate location and 
the numerical minimization for refinement. 
As illustrated in Fig.~\ref{fig1}, a map can be drawn 
for the zeros of the real and imaginary parts of the partition function 
$Z(\beta)$ in the plane of the complex inverse temperature 
$\beta \equiv \beta_\mathrm{R} + i\beta_\mathrm{I}$.
This can be done by using a one-dimensional root finder for 
$\beta_\mathrm{I}$ at a given $\beta_\mathrm{R}$.
Given the map in the wide range of $\beta_\mathrm{R}$, 
one can identify an approximate location of the crossing 
where the zeros of the real and imaginary parts meet each other.
The leading Fisher zero $\beta_1$ is given by the crossing
with the smallest magnitude of $\beta_\mathrm{I}$.
Starting from the approximate location of the crossing, 
$\beta_1$ is refined by numerically minimizing $|Z(\beta)|$.

While this two-step approach can work in principle with 
any normalization of the partition function, the WL estimate may 
prefer the particular form with 
\begin{equation} \label{eq:Z}
    \tilde{Z}(\beta) \equiv \frac{Z(\beta)}{Z(\beta_\mathrm{R})} 
    = \sum_{E}P(E;\beta_\mathrm{R})e^{-i\beta_\mathrm{I}E} 
    = \left\langle e^{-i\beta_\mathrm{I} E} 
    \right\rangle_{\beta_\mathrm{R}}
\end{equation}
because once the density of states $g(E)$ is obtained, 
the energy distribution $P(E;\beta_\mathrm{R})$ is computed 
straightforwardly at any real inverse temperature $\beta_\mathrm{R}$ as
\begin{equation} \label{eq:PE}
    P(E;\beta_\mathrm{R}) \equiv 
    \frac{1}{Z(\beta_\mathrm{R})}
    g(E) e^{-\beta_\mathrm{R}E} .
\end{equation}
The energy distribution is not affected by an arbitrary normalization 
factor for the WL estimate of $g(E) \equiv g_\mathrm{WL}(E)$ 
since it is canceled out with $Z(\beta_\mathrm{R})$.
The conventional MC simulations use a similar reweighting strategy 
in the grids of $\beta_R$.
In the HOTRG calculations, the normalization 
is not relevant since it gives the direct computation of $\ln Z(\beta)$, 
but we will still consider $\tilde{Z}(\beta)$ for comparison with 
the WL estimates.

The other method to compute the partition function zeros 
is to use a polynomial solver, while it is applicable only when 
the density of states $g(E)$ is prepared as a function of equally spaced 
energy $E \equiv E_n$. For a given $E_n = n \epsilon + \epsilon_0$, 
where $n$ is a non-negative integer, 
one can find the zeros of the partition function by solving 
a complex polynomial equation,
\begin{equation}
    Z(z) = e^{-\beta\epsilon_0} \sum_{n=0}^{n_\mathrm{max}} g_n z^n = 0 , 
\end{equation}
where $z \equiv \exp(-\beta\epsilon)$ and $g_n \equiv g(E_n)$.
The zeros can be computed for instance by 
using the \textsc{MPSolve} package~\cite{mpsolve}.
This approach is applicable to the case of the Ising, Potts, and 
six-state clock models where the energy is regularly spaced. 
However, it cannot be used for the models with continuous symmetry 
like the $XY$ model or the ones with irregular energies like 
the five-state clock model without introducing an artificial 
binning error. For our purpose of discussing the vulnerability 
due to the MC noises in the zero finder, the two-step approach
is also more intuitive. The leading zeros in this work 
are identified by using the two-step method.

\subsection{Wang-Landau sampling method}

The WL sampling method is used to estimate $g(E)$ in the Ising and Potts models.
The standard algorithm~\cite{Wang2001a,Wang2001b} is employed
with the stopping criterion of the modification factor at $10^{-8}$.
The histogram flatness criterion is set to be $0.99$ 
for the system sizes up to $L = 32$ and $0.95$ for the larger systems. 

The normalized partition function $\tilde{Z}(\beta)$ is evaluated 
based on a set of the WL samples of the density of states.
The energy probability distribution in Eq.~\eqref{eq:PE} is averaged 
over $N_s$ different WL samples of the density of states 
$\{g^{(k)}_\mathrm{WL}(E)\}$ as 
\begin{equation} \label{eq:PE_WL}
    P(E;\beta_R) \approx
    \bar{P}_\mathrm{WL}(E;\beta_R) = 
    \frac{1}{N_s}\sum_{k=1}^{N_s}
    \frac{g^{(k)}_\mathrm{WL}(E)e^{-\beta_R E}}
    {\sum_{E'} g^{(k)}_\mathrm{WL}(E')e^{-\beta_R E'} } .
\end{equation}
We have $N_s = 30$ from $30$ independent runs of the WL 
simulations for the model Hamiltonian examined.
The measurement uncertainty of $\tilde{Z}(\beta)$ 
and the location of the leading zero are computed
for the confidence level of 95\%
by repeating the bootstrap resampling processes
$1000$ times with the prepared WL samples of the density of states.

On the other hand, in the $p$-state clock model, while we mainly use 
the HOTRG method in this work, the WL method was used for the Fisher 
zero problem in the previous studies~\cite{Hwang2009,Kim2017}. 
However, the previous work~\cite{Kim2017} reported the strong limitation 
in accessible system sizes and argued that it was due to a fundamental 
property of the Fisher zero at the BKT transition rather than the WL algorithm itself.
In fact, for the six-state clock model, the WL density of states can be 
obtained for sizes up to $L=128$ within the same criterion used for the Ising model.
Even larger systems can be considered by using the parallel 
algorithm~\cite{Vogel2013,Vogel2014}.
While the five-state clock model needs the two-parameter
representation~\cite{Kim2017}, advanced strategies for acceleration
have been suggested recently for simulations in large systems
\cite{Valentim2015,Ren2016,Chan2017a,Chan2017b}.

\subsection{Higher-order tensor renormalization group}

The HOTRG method~\cite{Xie2012} provides a deterministic way of
evaluating the partition function in the tensor-network 
representation. For a classical spin model with local interactions 
in the square lattices, the partition function can be written as 
\begin{equation}
    Z(\beta) = \mathrm{Tr} \exp (-\beta H) 
    = \mathrm{Tr} \prod_i T_{x_i x^\prime_i y_i y^\prime_i} ,
\end{equation}
where $T_{x_i x^\prime_i y_i y^\prime_i}$ represents a local tensor 
with the indices of four legs associated with the bonds in the $x$ and 
$y$ directions. The complexity in this product of the local tensors 
can be truncated systematically in a controlled way by using 
the real-space renormalization group procedures~\cite{Levin2007,Xie2009,Zhao2010,Xie2012}.
In particular, the HOTRG method has been extended to the study 
of the Fisher~\cite{Denbleyker2014} and Lee-Yang~\cite{GarciaSaez2015}
zeros of the partition function evaluated at complex 
temperatures and fields.

Let us briefly review the HOTRG procedures. 
Initially, the model-dependent local tensor $T^{(0)}$ is prepared 
at each site, and then it is coarse-grained with the tensor 
in a neighboring site sharing a bond. 
In $2^N \times 2^N$ lattices, with translational invariance 
being assumed,
it takes $2N$ operations of the contraction applied alternatively along
the $x$ and $y$ directions to obtain the final coarse-grained tensor.
For instance, the $n$th operation of the contraction 
along the $y$ direction starts by computing
\begin{equation}
    M_{xx'yy'}^{(n)} = \sum_i T_{x_1 x_1^\prime y i}^{(n)} 
                       T_{x_2 x_2^\prime i y^\prime}^{(n)} ,
\end{equation}
where $x = x_1 \otimes x_2$ and $x^\prime = x_1^{\prime} \otimes x_2^\prime$. 
If the dimension of each leg of $T^{(n)}$ is $D$, then the dimension of $x$ and $x'$
increases to $D^2$. The dimension of the legs will be exponentially 
large as the contraction goes on if $M_{xx'yy'}$ is given directly 
to the new tensor. This is prevented by introducing the cutoff dimension 
$D_c$ for the spectral truncation in the singular value decomposition. 
The truncation error is controlled by increasing $D_c$.
The new coarse-grained tensor $T^{(n+1)}$ is then written as
\begin{equation}
    T_{xx'yy'}^{(n+1)} = \sum_{ij} U_{ix} M_{ijyy'}^{(n)}  U_{jx'}^{*} .
\end{equation}
For a real $\beta$, a real unitary matrix $U$ can be obtained by 
solving the eigenproblem of the semi-positive-definite matrix $AA^\dagger$ 
with $A_{x,x'yy'}=M_{xx'yy'}$ or $A_{x',xyy'}=M_{xx'yy'}$, where
the eigenvectors corresponding to the first $D_c$ largest eigenvalues
are only taken to determine $U$.  
For a complex $\beta$, we follow the strategy proposed in the previous 
work for the $XY$ model~\cite{Denbleyker2014} to find the orthogonal 
transformation with a real unitary matrix $U$. 
The previous work proposed to replace $AA^\dagger$ with 
$\mathrm{Re}[AA^\dagger]$, $\mathrm{Re}[AA^T]$, 
$\mathrm{Im}[AA^\dagger]$, or $\mathrm{Im}[AA^T]$. 
Here we choose $\mathrm{Re}[AA^\dagger]$ to preserve the trace 
of $AA^\dagger$ in our implementation of the HOTRG method. 
At every contraction, trying both of $A_{x,x'yy'}=M_{xx'yy'}$ 
and $A_{x',xyy'}=M_{xx'yy'}$ to build $AA^\dagger$, we pick the one 
with the smaller residue of the trace that is due to 
small eigenvalues not included in the largest $D_c$ eigenvalues 
constructing the new coarse-grained tensor $T^{(n+1)}$.

As the contractions are repeated, the components of $T^{(n)}$ can
increase to very large numbers. To avoid the numerical overflow, 
we normalize $T^{(n)}$ by a factor of $\lambda_n$ which we set 
to be the Euclidean norm of $T^{(n)}$. After $2N$ steps of 
the contractions, the partition function is evaluated as
\begin{eqnarray}
    \ln Z = \ln \mathrm{Tr}\,\tilde{T}^{(2N)}
    + \sum_{i=1}^{2N}{2^{2N-i}}\ln{\lambda_i} ,
\end{eqnarray}
where $\tilde{T}^{(n)} \equiv T^{(n)}/\lambda_n$ is the normalized tensor.

\begin{figure}[t]
\centering\includegraphics[width=0.48\textwidth]{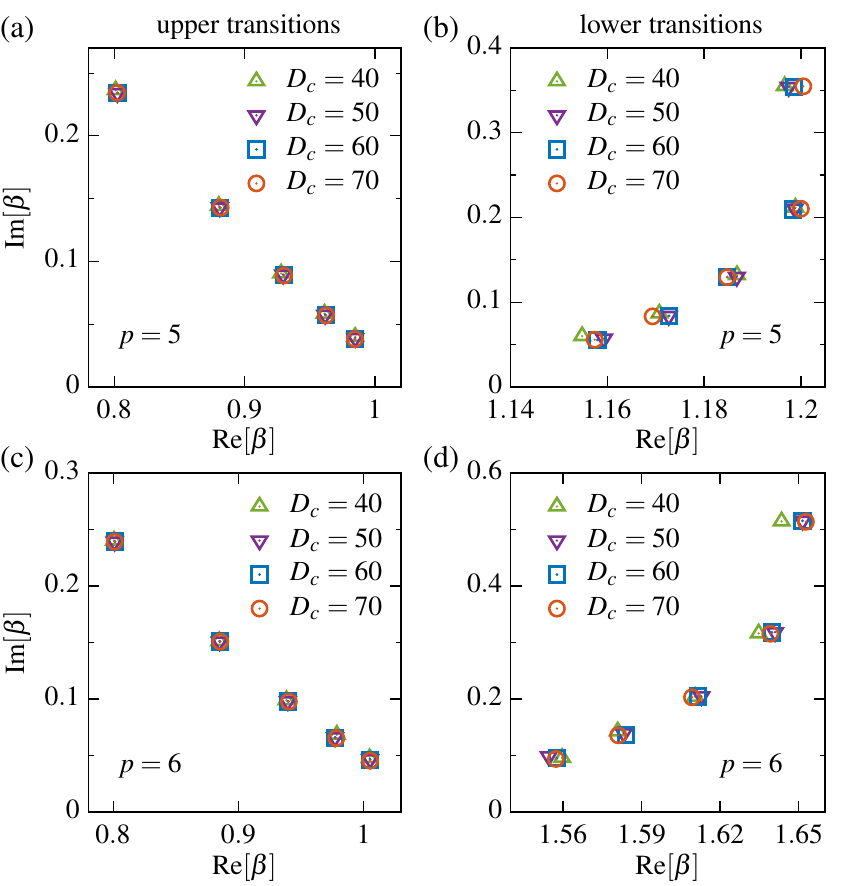}
\caption{\label{fig2}
Leading zero identification compared between different values of
the dimensional cutoff $D_c$ in the HOTRG procedures 
for the upper and lower transitions in the $p$-state clock model. 
The data points are available for the system sizes of $L=8,16,32,64,128$,
where the smaller $L$ corresponds to the larger value of $\beta_I$.
}
\end{figure}

The initial local tensor $T^{(0)} \equiv T$ depends on the model Hamiltonian.
For the Ising and $q$-state Potts model, 
the exact expression of $T$ is known~\cite{Zhao2010,Xie2012,Wang2014}. 
For the $p$-state clock model, we construct $T$ by using the same expansion 
technique previously used for the $XY$ model~\cite{Yu2014}.
The partition function $Z(\beta)$ of the $p$-state clock model is written as 
\begin{equation}
    Z(\beta) = \prod_{i} \sum_{\theta_i} \exp 
    \Big[ \beta \sum_{\langle i,j\rangle} \cos(\theta_i-\theta_j) \Big] 
    = \mathrm{Tr} \prod_i T_{x_i x^\prime_i y_i y^\prime_i} .
\end{equation}
The Boltzmann factor can be expanded by using
\begin{equation}
    e^{\beta\cos{\theta}} =
    \sum^\infty_{n=-\infty} I_n(\beta) e^{in\theta} \, ,
\end{equation}
where $I_{n}(\beta)$ is the modified Bessel function of the first kind. 
Summing out the spin angle variable $\theta_i$, we obtain
\begin{equation}\label{clocktensor}
    T_{xx'yy'} = 
    \sqrt[]{I_{x}(\beta)I_{x'}(\beta)I_{y}(\beta)I_{y'}(\beta)} 
    \delta_{\mathrm{mod}(x+y-x'-y',p),0} .
\end{equation}
The difference from the $XY$ model is indicated by 
$\mathrm{mod}(x+y-x'-y',p)$, which is $x+y-x'-y'$ modulo $p$, 
appearing because of the discrete values of $\theta$. 
Note that the initial tensor can be prepared alternatively 
by using the singular value decomposition~\cite{Chen2017,Chen2018}. 
While in the initial tensor, the magnitude of the coefficient 
$I_{n}(\beta)$ decays exponentially with increasing $n$, 
the cutoff dimension $D_c$ should be tested numerically 
for the required accuracy of final results.
The $XY$ model was examined previously with $D_c=40$ and
$50$~\cite{Yu2014,Denbleyker2014}. 
We increase the cutoff up to $D_c=70$, which is the largest $D_c$
available within our limitation of computational memory, 
for the identification of the leading zeros in the systems with 
sizes up to $L=128$~\cite{supp}.
Figure~\ref{fig2} displays the estimates with different $D_c$'s,
indicating that $D_c=60$ and $70$ are very similar. 
We use the dataset of $D_c=70$ for the FSS analysis 
presented in later sections.

\begin{figure}[t]
\centering\includegraphics[width=0.45\textwidth]{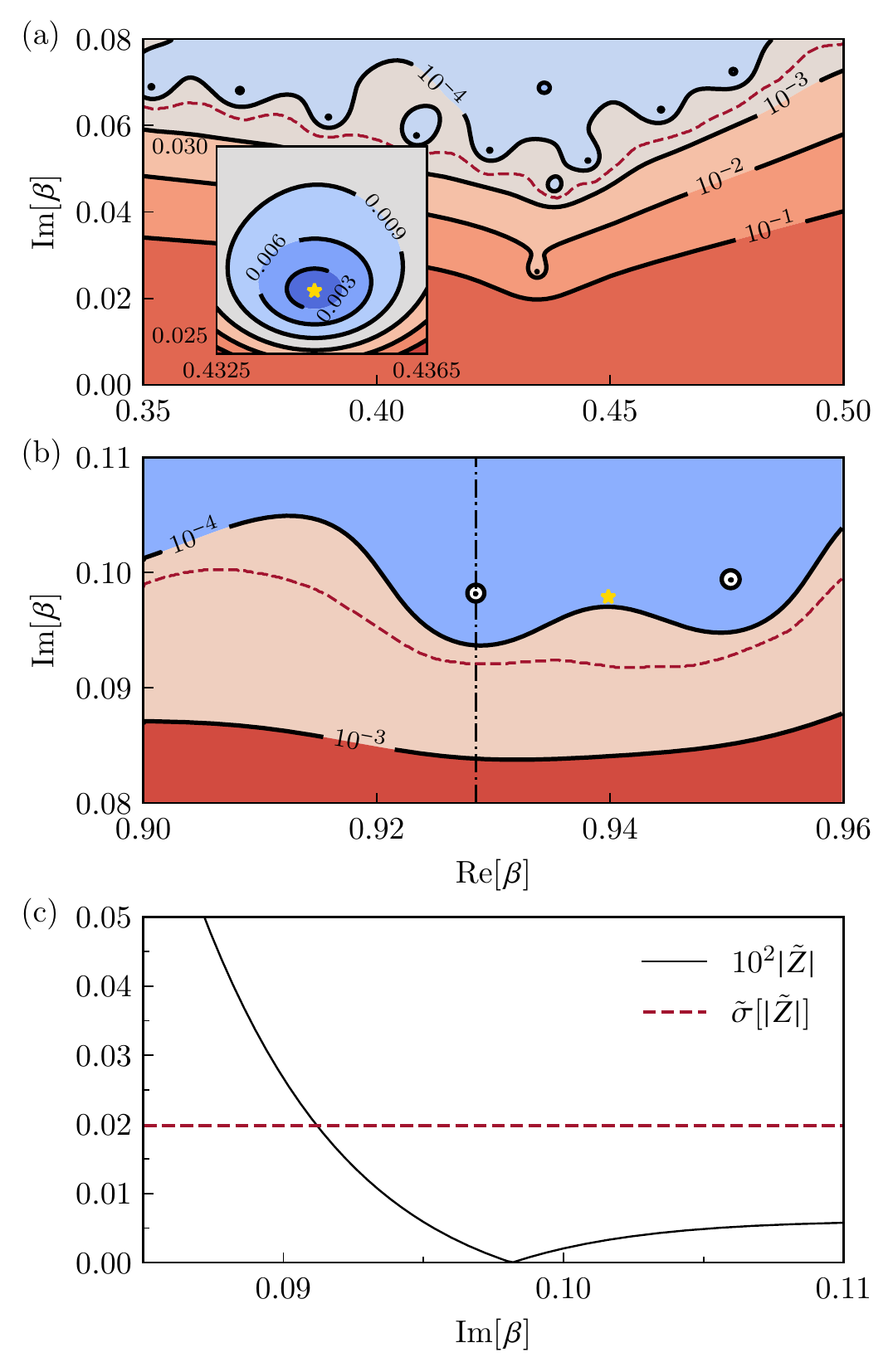}
\caption{\label{fig3}
Landscape of the normalized partition function $|\tilde{Z}|$
evaluated with the WL density of states for 
(a) the Ising~\cite{supp} and 
(b) six-state clock models~\cite{Kim2017}
in the square lattices of $L=32$. 
The dashed lines in (a) and (b) indicate the boundary 
above which $|\tilde{Z}|$ is smaller than the uncertainty 
estimate $\tilde{\sigma}[|\tilde{Z}|]$. 
The location of the true leading zero obtained from 
the HOTRG calculations is marked with the filled star.
(c) $|\tilde{Z}|$ and $\tilde{\sigma}[|\tilde{Z}|]$ plotted along 
the dash-dotted line in (b). 
}
\end{figure}

\section{Uncertainty of finding the leading zero under stochastic noises}
\label{sec:error}

The presence of stochastic noises is a general property of any MC estimator.
The questions that we address in this section are how much one 
can trust the identification of the leading Fisher zero under 
the stochastic errors of the partition function estimate and 
how it depends on a particular character of the phase transition. 
These questions were briefly considered by one of us 
in the previous work~\cite{Kim2017} which conjectured that
based on the Gaussian approximation, the numerical tolerance 
to the noises is related to the critical behavior of the specific heat. 
We examine this conjecture beyond the Gaussian approximation by providing 
more detailed analysis with demonstrations in the Ising, Potts, and clock models.

\subsection{System-size scaling of the uncertainty criterion}

Figure~\ref{fig3} presents examples of reliable and unreliable 
identifications of the leading zeros in the presence of the stochastic
uncertainty of the partition function estimate.
Finding the zeros of the normalized partition function $\tilde{Z}(\beta)$ 
under the uncertainty should accompany with a test for 
a signal-to-noise ratio to lend confidence to the search for the zero. 
The minimal condition can be set for the ``hill'' of $|\tilde{Z}|$ 
surrounding the location of the zero to be higher than the uncertainty 
level [see the dashed line in Fig.~\ref{fig3}(a)]; otherwise, 
the ``valley'' of $|\tilde{Z}(\beta)|=0$ is untrusted as indicated 
in Fig.~\ref{fig3}(c). 
The boundary of confidence~\cite{Alves1992,Denbleyker2007,Denbleyker2014}
can be given by the measurements based on the WL dataset 
as a line above which the magnitude of $|\tilde{Z}(\beta)|$ is 
smaller than its uncertainty measure $\tilde{\sigma}[|\tilde{Z}(\beta)|]$. 

While the boundary of confidence can be computed in MC simulations, 
one can build useful intuition about how the boundary would evolve 
with increasing system sizes from the analytic approach based on 
the Gaussian approximation of the energy distribution. 
In Ref.~\cite{Alves1992}, the random sampling with the Gaussian energy
distribution provided the standard error 
$\tilde{\sigma}[|\tilde{Z}(\beta)|] = [(1-|\tilde{Z}|^2)/n_s]^{1/2}$,
which defined the radius of confidence from the real axis as 
\begin{equation} \label{eq:radius}
    R = \sqrt{\ln (n_s+1)}/\sigma_E ,
\end{equation}
where $n_s$ is the size of the samples,
and $\sigma_E^2 \equiv \langle E^2 \rangle - \langle E\rangle^2$ 
is the variance of energy at a given $\beta_\mathrm{R}$. 
The extension to the quasi-Gaussian distribution was also discussed 
in Ref.~\cite{Denbleyker2007}. Despite the difference from realistic
energy distributions, Eq.~\eqref{eq:radius} still provides 
an important implication on the numerical accessibility to 
the leading zero that turns out to differ with the type of 
the associated phase transition.

The system-size dependence of $R$ is encoded in the energy variance 
that is proportional to the heat capacity.
Since the heat capacity is extensive, one may anticipate that 
$R \sim L^{-d/2}$ in the $d$-dimensional lattices, 
indicating that the area where we can trust the estimate 
of $\tilde{Z}$ shrinks with a power law as $L$ increases.
The decrease of $R$ is even faster in the vicinity of the leading 
zero along the line of $\beta_\mathrm{R} = \mathrm{Re}[\beta_1]$ 
because it corresponds to a pseudotransition point at which 
the specific heat $c_L^*$ becomes critical in the ordinary phase transition. 
However, an important missing piece in this argument is that 
the leading zero, which we want to identify, is also moving 
toward the real axis as the system size increases.

Therefore, what we need to consider is the race between $R(c_L^*)$ 
and $\mathrm{Im}[\beta_1]$, both of which decrease 
with increasing $L$. If $R$ is always larger than $\mathrm{Im}[\beta_1]$
regardless of $L$, then one can successfully locate the leading zero 
even at a very large system within the uncertainty of the MC estimate. 
If $R$ becomes smaller than $\mathrm{Im}[\beta_1]$ at some point of $L$, 
then the zero identified under the noises is likely to be accidental and 
thus hardly trusted. Because $c_L^*$ and $\mathrm{Im}[\beta_1]$ 
are both governed by the critical behaviors, one may reach an intuition
that while the FSS behavior of the leading zero characterizes
the phase transition, the character of the transition may also influence 
reversely the numerical difficulty of finding the leading zero.

In the first-order transitions, the diverging specific heat 
$c_L^* \sim L^d$ at a pseudotransition point 
leads to $R \sim L^{-d}$ which coincides with  
the expected behavior of $\mathrm{Im}[\beta_1] \sim L^{-d}$.  
In the second-order transitions with the critical exponent 
$\alpha > 0$, the specific heat $c_L^* \sim L^{\alpha/\nu}$
leads to $R \sim L^{-(d\nu+\alpha)/2\nu}$ that becomes $R\sim L^{-1/\nu}$ 
with the hyperscaling relation $d\nu = 2 - \alpha$, and the leading zero
has the same scaling behavior of $\mathrm{Im}[\beta_1]\sim L^{-1/\nu}$. 
Therefore, the ordinary first-order and second-order transitions exhibit 
$R \sim \mathrm{Im}{\beta_1}$ regardless of $L$, suggesting that 
the leading zero may be marginally accessible even at a very large 
system under finite stochastic uncertainty of $|\tilde{Z}|$. 

On the other hand, the situations are very different in the $XY$ model 
where the specific heat does not diverge
\cite{Kosterlitz1973,Kenna1995,Kenna1997,Kenna2006}. 
The radius $R \sim L^{-d/2}$ decreases much faster than the imaginary part 
of the leading zero that is expected to scale as
$\mathrm{Im}[\beta_1] \sim [\ln(bL)]^{-\tilde{q}}$
with $\tilde{q} = 1+1/\nu$ at a very large $L$~\cite{Denbleyker2014}.
The singular part of the specific heat with the logarithmic 
correction~\cite{Kenna1995,Kenna1997,Kenna2006} is proportional to
$L^{-d}(\ln L)^{2\tilde{q}}$ which is comparable to $\mathrm{Im}[\beta_1]^{-2}$;
however, the main contribution to $R$ comes from 
the constant regular part since the singular part quickly decreases 
with increasing $L$.
This leads to $R \ll \mathrm{Im}[\beta_1]$ at a large $L$, 
implying that for the BKT transitions, a reliable identification 
of the leading zero is fundamentally limited to small systems 
within the MC estimate of $|\tilde{Z}|$.

While Eq.~\eqref{eq:radius} indicates a connection between
the numerical feasibility of finding the leading zero and the critical phenomena, 
MC simulation are often performed to keep the measurement error at 
a certain level. Thus, in practice, it is meaningful to consider 
the criterion for a fixed uncertainty $\tilde{\sigma}_0$,
\begin{equation}
    |\tilde{Z}(\beta)| \ge \tilde{\sigma}_0 ,
\end{equation}
which examines whether the hill of $|\tilde{Z}|$ surrounding 
the leading zero is visible above the uncertainty level~\cite{Kim2017}. 
For the Gaussian energy distribution, $|\tilde{Z}(\beta)|$ 
at a complex value of $\beta = \beta_\mathrm{R} + i \beta_\mathrm{I}$ 
is calculated as
\begin{equation} \label{eq:Zgau}
    |\tilde{Z}(\beta)| = \exp\left[-L^d c_L^*
    \frac{\beta_\mathrm{I}^2}{2\beta_\mathrm{R}^2}\right] .
\end{equation}
At the location of the leading zero, $\beta = \beta_1$, 
the leading-order behavior of $|\tilde{Z}(\beta_1)| \equiv Q(L)$ 
for a large $L$ can be written as
\begin{eqnarray}
    Q_\mathrm{1st}(L) &=& A_L\exp[-a_0 L^{-d}] , \label{eq:Q_1st} \\
    Q_\mathrm{2nd}(L) &=& A_L\exp[-a_0 L^{-\alpha/\nu}] , \label{eq:Q_2nd} \\
    Q_\mathrm{BKT}(L) &=& A_L\exp[-a_0 L^d (\ln bL)^{-2\tilde{q}}] , \label{eq:Q_BKT}
\end{eqnarray}
for the first-order, second-order, and BKT transitions, respectively, 
where a constant $a_0$ is given by the regular part of $c_L^*$.
The singular part of $c_L^*$ is canceled out together with
$\mathrm{Im}[\beta_1]$ as discussed for $R$. It contributes to
to the factor $A_L$, but it may also contain the logarithmic or 
next-to-leading order corrections of $c_L^*$.
For instance, in the Ising model, the specific heat with $\alpha=0$ 
diverges logarithmically as $c_L^* \sim \ln L$, leading to
the power-law decay of $A_L$ as
\begin{equation} \label{eq:Q_Ising}
    Q_\mathrm{Ising}(L) = A_0 L^{-a} . 
\end{equation}
For a small $\alpha$, considered the next-to-leading order correction
term as in $c_L^* \sim L^{\alpha/\nu}(1 - a_\omega L^{-\omega})$,
we may rewrite $Q(\beta_1)$ as
\begin{equation} \label{eq:Q_weak}
    Q_\mathrm{2nd}(L) = A_0 \exp[-a_0L^{-\alpha/\nu} 
    + a_1 L^{-\omega}] ,
\end{equation}
indicating that $Q_\mathrm{2nd}(L)$ can decrease slowly as
$L$ increases when $a_0 < 0$, or $a_1L^{-\omega}$ is positive and dominant.

Therefore, for the first-order transition, $Q_\mathrm{1st}$ is 
an increasing function of $L$, implying that the same uncertainty level
is enough for large systems. While it works similarly in the second-order 
transition, one may need more accurate estimates at a larger system
if $\alpha$ is small. For the BKT transitions, $Q_\mathrm{BKT}$ 
exponentially decays with increasing $L$, implying that a larger system
requires exponentially more accurate estimates that may not be feasible 
with usual MC simulations.

\begin{figure*}[t]
\centering\includegraphics[width=\textwidth]{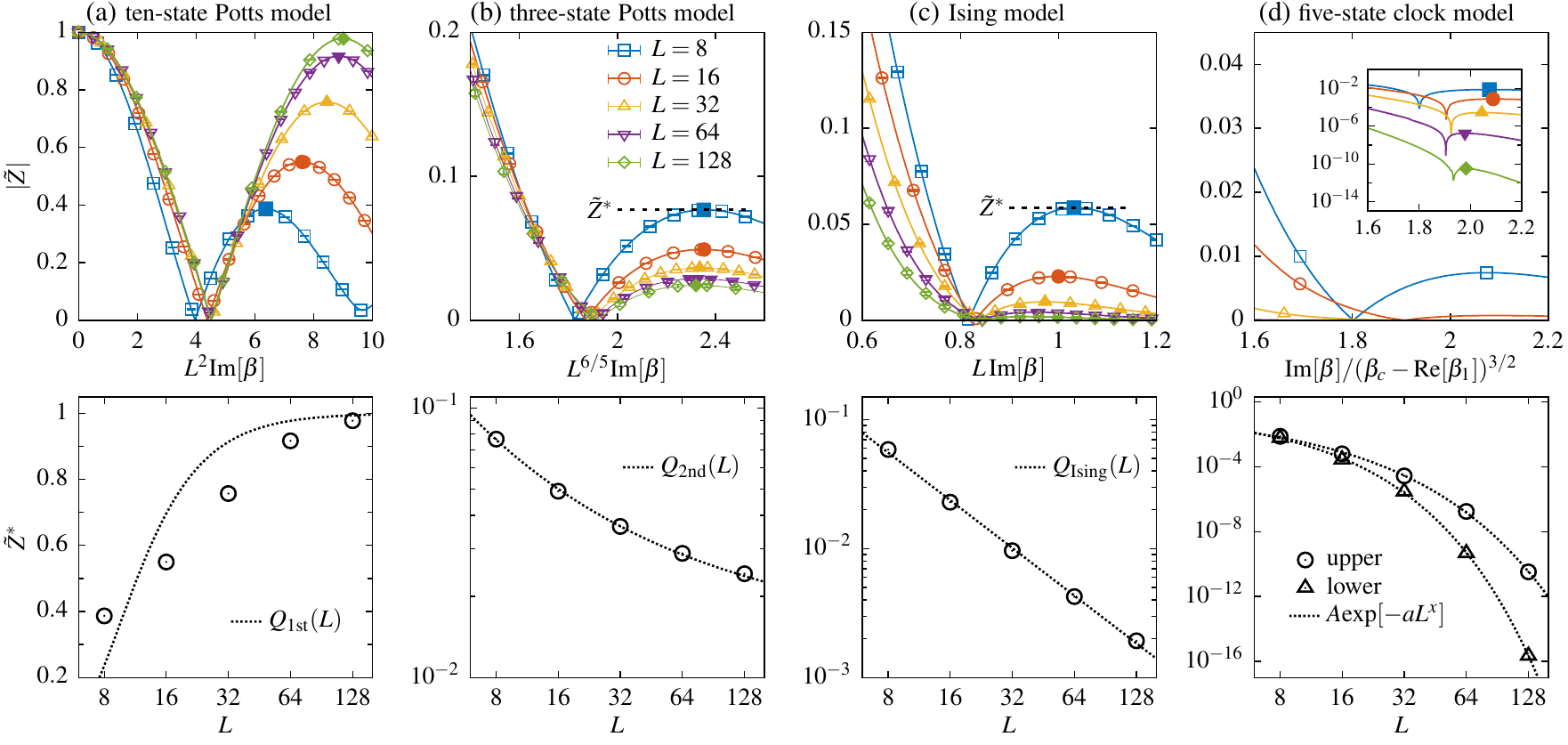}
\caption{\label{fig4}
Numerical visibility of the leading zeros
associated with the different types of phase transition.
The normalized partition function $\tilde{Z}$ is plotted 
in the vicinity of the leading zero for (a) 10-state Potts, 
(b) three-state Potts, (c) Ising, and (d) five-state clock models.
The hill height $\tilde{Z}^*$, the maximum of $|\tilde{Z}|$ 
above the leading zero in the imaginary axis is marked 
with a filled symbol. In the bottom panels, the system-size scaling
behavior of $\tilde{Z}^*$ is compared with a curve fit to $Q(L)$ 
predicted from the Gaussian approximation. The partition functions 
shown here are obtained by using the WL density of states for the Ising 
and Potts models and by using the HOTRG method with $D_c=70$ 
for the five-state model. The identified location
of the leading Fisher zeros are tabulated in Supplemental Material~\cite{supp}.}
\end{figure*}

\subsection{Numerical visibility of the leading zeros in spin models}

A caveat of the above argument is that the Gaussian form of 
the energy distribution becomes a crude approximation
near the actual location of the zero. It is well known that 
the Gaussian energy distribution cannot produce $Z=0$
as also indicated in Eq.~\eqref{eq:Zgau}. The previous work
\cite{Kim2017} argued that Eq.~\eqref{eq:Zgau} works as an envelope 
function of $\tilde{Z}(\beta)$, and thus the analytic connection 
between the numerical difficulty and the critical behavior 
of the specific heat can be still intuitive. 
Therefore, it is still important to check numerically, 
in the realistic spin models, how the maximally
tolerable uncertainty for a trusted identification of the zero 
scales with the system size.

As illustrated in Fig.~\ref{fig3}, the uncertainty level that can 
still reveal the valley of $|\tilde{Z}|=0$ is bound by the height of 
the hill surrounding the location of the zero.
Thus, we compare the behavior of the hill height just above 
the leading zero, which we denote by $\tilde{Z}^*$, with the predicted 
scaling behavior of $Q(L)$.
Figure~\ref{fig4} presents the system-size dependence of $\tilde{Z}^*$ 
in the Potts, Ising, and clock models to examine the different types 
of phase transition.
The partition functions in the Potts and Ising models are evaluated
based on the WL density of states, and the calculations are done 
in the clock model by using the HOTRG method.
It turns out that despite the quantitative difference from $Q(L)$,
the FSS behaviors of $\tilde{Z}^*(L)$ agree well with the behavior 
of  $Q(L)$ predicted based on the Gaussian approximation of 
the energy distribution. 

In the 10-state Potts model undergoing the first-order transition,
it is notable that the estimate of $\tilde{Z}^*$ 
increases toward unity as $L$ increases [see Fig.~\ref{fig4}(a)]. 
While the numerical data of $\tilde{Z}^*$ does not fit precisely
to the line of $Q_\mathrm{1st}(L)$ in Eq.~\eqref{eq:Q_1st}, 
the prediction of the increasing behavior is essentially valid.
The excellent contrast between the hill and valley 
of $\tilde{Z}^*$ guarantees that the location of the leading zero
can be accessible under finite noises even at a large system. 
The comparison with other phase transitions discussed below 
suggests that within the MC simulations, the numerical identification 
of the leading zero is the most stable at the first-order transition.

The examples of the second-order transitions also show excellent 
agreement with the expectation from the Gaussian approximation.
The three-state Potts model shown in Fig.~\ref{fig4}(b) presents
that the hill height $\tilde{Z}^*$ decreases with $L$ but tends to 
asymptotically converges. The data points of $\tilde{Z}^*$ show
a very good curve fit to $A\exp[-a_0 L^{-\alpha/\nu}]$ 
with $a_0<0$ and $\alpha/\nu \approx 0.4$ given in Eq.~\eqref{eq:Q_weak}. 
The parameters are consistent with the conjectured value of 
$\alpha/\nu = 2/5$~\cite{denNijs1979}
and the previous FSS test of the specific heat maximum where 
the negative constant term ($a_0 < 0$) was indicated~\cite{Nagai2013}.
In the Ising model presented in Fig.~\ref{fig4}(c), the data points
$\tilde{Z}^*$ indicate a power-law decrease as expected 
from $Q_\mathrm{Ising}(L)$ in Eq.~\eqref{eq:Q_Ising}.
While the decreasing behavior suggests that the stochastic error 
should decrease accordingly to identify the leading zero, 
the measured uncertainty of our WL estimates is well below 
the hill level of $\tilde{Z}^*$ in the tested range of the system sizes.

On the other hand, in the five-state clock model, we observe 
an exponential decay of $\tilde{Z}^* \sim \exp(-aL^x)$ 
in the HOTRG calculations of the partition function 
[see Fig.~\ref{fig4}(d)]. 
This indicates that if there were finite noises, then the valley-hill 
structure around the leading zero would get exponentially less visible 
with increasing system size. Although the observed scaling behavior of
$\tilde{Z}^*$ is different from Eq.~\eqref{eq:Q_BKT}, 
both reach the same conclusion that the search for the leading zero 
would become extremely vulnerable against stochastic noises, 
implying that the MC methods are inadequate to the Fisher-zero 
study of the BKT transitions. This emphasizes the advantage
of HOTRG as a deterministic method whose accuracy is 
controlled with the cutoff dimension and free from stochastic noises.

In the next section, we present the FSS analysis with the leading zeros 
identified in the HOTRG calculations to characterize the BKT features 
of the upper and lower transitions in the five- and six-state clock models.

\section{Two BKT transitions in the clock model}
\label{sec:clock}

Let us begin this section by summarizing the problems that remain
unsolved in the previous Fisher-zero study of the $p$-state clock 
model~\cite{Kim2017}. First, for $p=5$, the transition point
suggested by the leading zeros seemed to deviate from the previous 
estimates when the BKT exponent $\nu=1/2$ is used, 
while $\nu=1/2$ has been verified in the phenomenological FSS
analysis~\cite{Borisenko2011} and the analysis of the helicity modulus
redefined for the discrete symmetry~\cite{Kumano2013,Chatelain2014}. 
Second, the leading zeros at the lower transitions showed 
an arclike FSS trajectory which was different from 
the power-law trajectory expected from the $XY$ model.
The leading zero behavior at the lower transition remains 
unexplained. 

An obvious criticism to the previous analysis based on the WL 
density of states in Ref.~\cite{Kim2017} was that the system sizes 
examined were too small to draw any conclusive results. 
Here we provide the FSS analysis with the leading zeros obtained 
by using the HOTRG calculations in the systems of sizes up to $L=128$.
Nevertheless, it turns out that the finite-size effects are still strong
so that the known leading-order ansatz is not enough to explain the observed
behaviors, suggesting that subleading order corrections 
are necessary to be included in the FSS analysis within 
the available system sizes.

\subsection{Logarithmic correction to the finite-size-scaling ansatz}

The FSS behavior of the leading Fisher zero in the previous study 
of the $XY$ model~\cite{Denbleyker2014} was derived by extending 
the pseudotransition temperature obtained from the system-size scaling 
of the correlation length $\xi_L(\beta) \propto L$ into the domain of 
the complex temperature. It was found that a complex pseudotransition 
temperature $\beta_L \equiv \beta_x(L) + i\beta_y(L)$ would 
behave as $\beta_y \propto (\beta_c - \beta_x)^{3/2}$ at a small $\beta_y$
at a large $L$. In the present study, we incorporate the logarithmic
finite-size correction into the correlation length, which turns 
out to be essential to the FSS analysis of the leading Fisher zeros.

\begin{figure}[t]
\centering\includegraphics[width=0.48\textwidth]{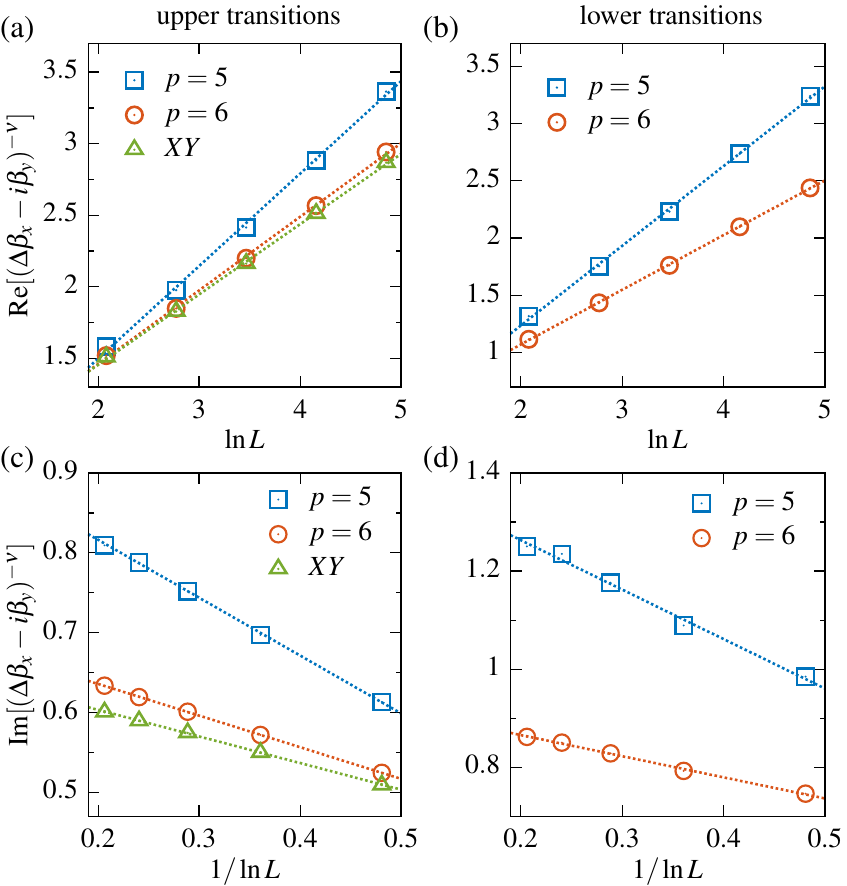}
\caption{\label{fig5}
Logarithmic finite-size corrections to the scaling ansatz.
The real and imaginary parts of $(\Delta\beta_x-i\beta_y)^{-\nu}$ 
are shown at the transitions in the $p$-state clock model 
and the $XY$ model. The postulated BKT exponent $\nu=1/2$ is used.
The dotted lines are given by the curve fits of the parameters 
in Eq.~\eqref{eq:complex_beta}. The numerical values of 
the parameters are listed in Ref.~\cite{supp}. 
The HOTRG data at $D_c=70$ are used.
}
\end{figure}

While the FSS ansatz of the correlation length is typically written 
as $\xi_L(\beta) / L = a_0$ with a constant $a_0$ for a large $L$,
the previous MC study of the second moment correlation length
in the $XY$ model~\cite{Hasenbusch2005} indicated the presence of 
the logarithmic correction. Assumed that it works in the same way
in the complex domain, we may begin with the ansatz written as
\begin{equation} \label{eq:xi}
   \frac{\xi_L(\beta)}{L} = a_0 + \frac{a_1}{\ln L} + O[(\ln L)^{-2}],
\end{equation}
where the constants $a_0$ and $a_1$ are complex numbers.
Given the BKT ansatz of $\xi = A\exp[1/(at^\nu)]$, we can write 
an equation for a reduced temperature $t\equiv t_L$ at a finite $L$ as
\begin{equation} \label{eq:t_L}
    \frac{1}{at_L^\nu} = z_0 + \ln L + \frac{z_1}{\ln L} + O[(\ln L)^{-2}],  
\end{equation}
where $z_0 = \ln (a_0/A)$, and $z_1 = a_1/a_0$.  
To the leading and next-to-leading orders for the real and 
imaginary parts of the right-hand side, the final FSS ansatz 
is written as
\begin{equation} \label{eq:complex_beta}
    (\Delta\beta_x \pm i\beta_y)^{-\nu} \simeq 
    a\ln bL + i\left( c_0 - \frac{c_1}{\ln L} \right),
\end{equation}
where the complex reduced temperature $t_L$ for a transition point 
$\beta_c$ is written as $t_L \equiv \Delta\beta_x \pm i\beta_y$ 
with $\Delta\beta_x = | \beta_c - \beta_x |$. The sign of 
$\pm \beta_y$ is irrelevant because of the symmetry of the Fisher zero,
and we choose $\beta_y>0$ and $t_L = \Delta\beta_x - i\beta_y$. 
The logarithmic behaviors of the real and imaginary parts are
evident in the numerical tests with $\nu=1/2$ in Fig.~\ref{fig5}.

Equation~\eqref{eq:complex_beta} can be solved for 
$\Delta\beta_x = r_L \cos \theta_L$ and $\beta_y = r_L \sin \theta_L$
in the polar coordinates of the complex inverse temperature. 
The radius $r_L$ and the angle $\theta_L$ are written as
\begin{eqnarray}
    r_L &=& (a\ln bL)^{-\frac{1}{\nu}}
    \left[ 1 + \psi_L^2 \right]^{-\frac{1}{2\nu}}, \\
    \theta_L &=& \frac{1}{\nu} \tan^{-1} \psi_L,
\end{eqnarray}
where we define the size-dependent parameter $\psi_L$ as 
\begin{equation} \label{eq:psi_L}
    \psi_L = \frac{1}{a\ln bL}\left[ c_0 - \frac{c_1}{\ln L} \right]
    \equiv \frac{\mathrm{Im}[(\Delta\beta_x - i\beta_y)^{-\nu}]}
    {\mathrm{Re}[(\Delta\beta_x - i\beta_y)^{-\nu}]}.
\end{equation}

For a small $\psi_L$, one can write $\Delta\beta_x$ and $\beta_y$ as
\begin{eqnarray}
    \Delta\beta_x &=& (a\ln bL)^{-\frac{1}{\nu}} 
    \left[ 1 - B_1 \psi_L^2 + O(\psi_L^4) \right], 
    \label{eq:fss_psi_beta_x}\\
    \beta_y &=& \frac{1}{\nu}(a\ln bL)^{-\frac{1}{\nu}} \psi_L
    \left[ 1 - B_2 \psi_L^2  + O(\psi_L^4) \right],
    \label{eq:fss_psi_beta_y}
\end{eqnarray}
where $B_1 = \frac{1}{2\nu} + \frac{1}{2\nu^2}$, and
$B_2 = \frac{1}{3} + \frac{1}{2\nu} + \frac{1}{6\nu^2}$. 
One can further expand these equations in powers of $1/\ln L$ as
\begin{eqnarray}
    \Delta\beta_x &\propto& (\ln bL)^{-\frac{1}{\nu}}
    \left[ 1 - B'_1 (\ln L)^{-2} + O[(\ln L)^{-3}] \right], 
    \label{eq:fss_beta_x} \\
    \beta_y &\propto& (\ln bL)^{-1-\frac{1}{\nu}}
    \left[ 1 - B'_2 (\ln L)^{-1} + O[(\ln L)^{-2}] \right], 
    \label{eq:fss_beta_y}
\end{eqnarray}
where $B'_1 = B_1 c_0^2 / a^2$, and $B'_2 = c_1/c_0$.
In the asymptotic limit, they approach the lines of
$\Delta\beta_x \propto (\ln bL)^{-1/\nu}$ and 
$\beta_y \propto (\ln bL)^{-1-1/\nu}$,
reproducing the simple power-law trajectory of 
$\beta_y \propto \Delta\beta_x^{1+\nu}$ 
that was proposed in Ref.~\cite{Denbleyker2014}. 
To take into account the logarithmic correction terms
in $\Delta\beta_x$ and $\beta_y$, 
the leading zero trajectory can be expressed as
\begin{equation}\label{eq:trajectory}
    \Delta\beta_x = w_1 \beta_y^{\frac{1}{1+\nu}} 
    + w_2 \beta_y + w_3 \beta_y^{2-\frac{1}{1+\nu}}
    + O\big(\beta_y^{3-\frac{2}{1+\nu}}\big) \, ,
\end{equation}
where the coefficients can be determined perturbatively
from the asymptotic solution.
While this expression includes the higher-order corrections, 
it is still unclear how the trajectory can bend 
like an arc as previously observed at the lower transitions 
in the $p$-state clock model~\cite{Kim2017}. 

\begin{figure}[t]
\centering\includegraphics[width=0.48\textwidth]{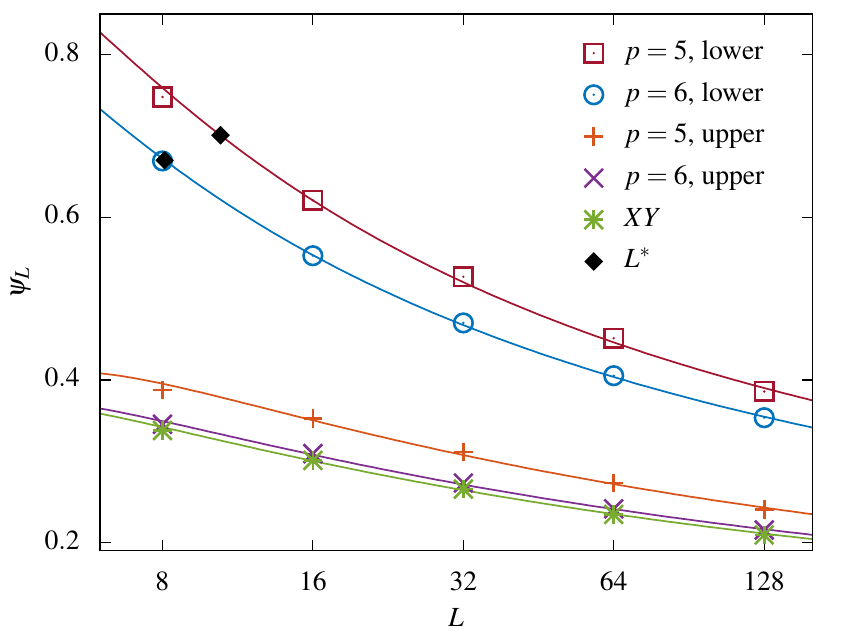}
\caption{\label{fig6}
System-size dependence of $\psi_L$ in the $p$-state clock model 
and the $XY$ model. The solid lines indicate Eq.~\eqref{eq:psi_L}
with the fitting parameters determined in Fig.~\ref{fig5}. 
The filled symbol denotes $L^*$ at which $\Delta\beta_x$ has maximum 
in Eq.~\eqref{eq:fss_exact_beta_x}. 
}
\end{figure}

We find that at the postulated BKT exponent $\nu=1/2$, 
the closed-form expressions of $\Delta\beta_x$ and $\beta_y$ 
are obtained to show more explicitly the character of 
the leading-zero trajectory at a finite $L$. 
At $\nu=1/2$, Eq.~\eqref{eq:complex_beta} provides the expressions,
\begin{eqnarray}
    \Delta\beta_x &=&  \frac{\psi_L^2(1-\psi_L^2)}{(1+\psi_L^2)^2}
    \left[ c_0 - \frac{c_1}{\ln L} \right]^{-2},
    \label{eq:fss_exact_beta_x} \\
    \beta_y &=& \frac{2\psi_L^3}{(1+\psi_L^2)^2}
    \left[ c_0 - \frac{c_1}{\ln L} \right]^{-2}.
    \label{eq:fss_exact_beta_y}
\end{eqnarray}
It turns out that as $\psi_L$ decreases, $\Delta\beta_x$ increases 
first at a large $\psi_L$ and then starts to decrease 
when $\psi_L$ becomes smaller than a certain value. 
This contrasts with the monotonic increase in $\beta_y$. 
Thus, if $\psi_L$ is not small, then one may find $L^*$ at which the slope 
of $\Delta\beta_x$ change its sign, leading to an arclike trajectory 
in the complex plane.
In the numerical tests shown in Fig.~\ref{fig6}, 
the leading zeros at the lower transitions have much larger values of
$\psi_L$ than at the upper transitions, explaining the strong
finite-size effects observed at the lower transitions.
Below we demonstrate the finite-size behaviors discussed in
this section by using the HOTRG data of the leading zeros.

\begin{figure}[t]
\centering\includegraphics[width=0.48\textwidth]{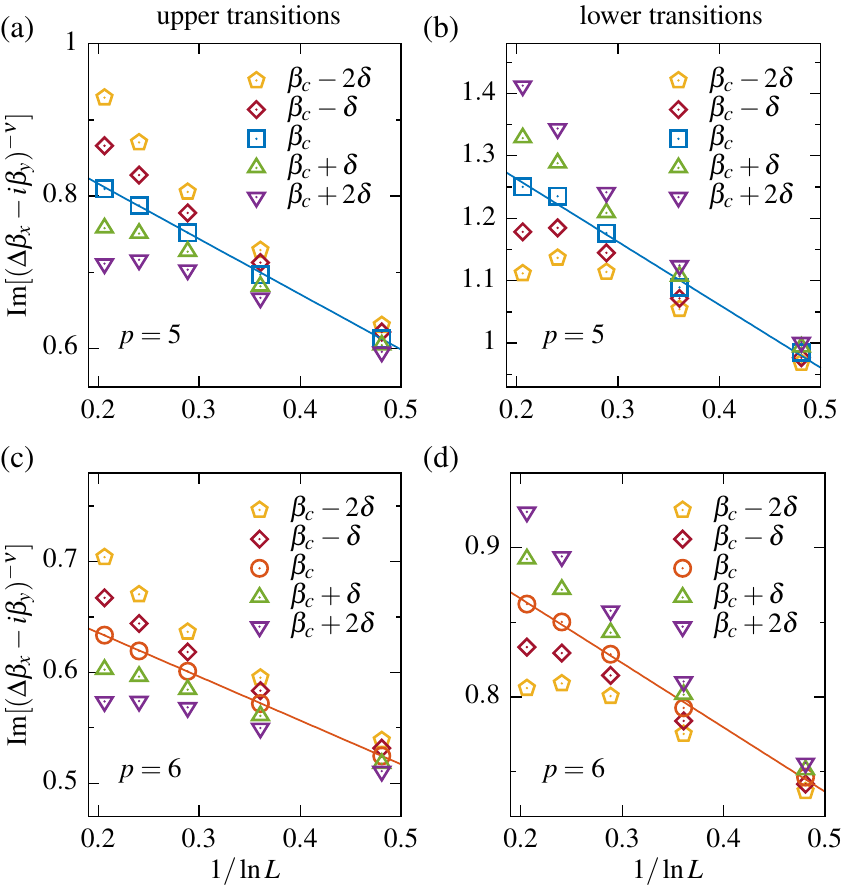}
\caption{\label{fig7}
Determination of the transition points. The ansatz with the $1/\ln L$  
correction is employed to locate the transition point $\beta_c$ 
at the upper [(a) and (c)] and lower [(b) and (d)] transitions 
in the $p$-state clock model. The sensitivity of the $1/\ln L$ 
behavior (solid line) is examined 
by moving away from $\beta_c$ with the step size $\delta = 0.004$. 
The HOTRG data and the estimates of $\beta_c$ at $D_c=70$ are used.
}
\end{figure}

\subsection{Finite-size-scaling behaviors of the leading Fisher zeros}

The logarithmic FSS behavior in the imaginary part of
Eq.~\eqref{eq:complex_beta} plays an essential role to determine 
the transition points from the leading Fisher zero data. While both of 
the real and imaginary parts expect the $\ln L$ dependence as shown 
in Fig.~\ref{fig5}, the imaginary part of 
$(\Delta\beta - i\beta_y)^{-\nu}$ 
responds much more sensitively to the change of $\beta_c$, providing 
a stable curve fit to locate $\beta_c$ in practice.
Figure~\ref{fig7} demonstrates the behavior of 
$\mathrm{Im}[(\Delta\beta - i\beta_y)^{-\nu}]$  
with the postulated BKT exponent of $\nu=1/2$, indicating
the systematic deviations of the data points from the straight line 
as it moves away from the determined value of $\beta_c$.

Table~\ref{tab:beta_c} lists our Fisher-zero estimates of $\beta_c$ 
based on the HOTRG data computed at $D_c=60$ and $70$ 
and the previous results based on various different measures 
at the upper and lower transitions in the $p$-state clock model 
for $p=5$ and $6$. 
Our estimates at both $D_c$'s are well in the range of 
the previous estimates.
The estimate of $\beta_c$ is obtained by solving the least-squares
problem to minimize the absolute difference between
$\mathrm{Im}[\Delta\beta_x - i\beta_y]^{-1/2}$ and $c_0 - c_1/\ln L$
in Eq.~\eqref{eq:complex_beta}. 
The HOTRG calculations are deterministic and 
provide a single set of the leading Fisher zero data 
at each $D_c$~\cite{supp}. 
The error given in the parentheses in Table~\ref{tab:beta_c}
is the fitting uncertainty at a given $D_c$ measured by 
the jackknife variance with one data point being discarded.  
The fitting uncertainty gets smaller with the larger $D_c$,  
supporting the $1/\ln L$ correction ansatz in 
Eq.~\eqref{eq:complex_beta}.
Comparing with the previous Fisher zero study 
using the WL method~\cite{Kim2017},
the reasons for the better agreement between the present results and 
the other estimates are twofold. First, our HOTRG dataset
covers up to $L=128$ which is much larger than $L\lesssim 32$ of 
the previous WL study. 
Second, while Ref.~\cite{Kim2017} relied on the form of the trajectory 
that is valid in the asymptotic limit, our method of locating $\beta_c$ 
benefits from the logarithmic finite-size correction in 
the next-to-leading order, providing a better access to 
finite systems at both of the upper and lower transitions.

The finite-size influence of neglecting the higher order logarithmic 
terms in the FSS ansatz of $(\Delta\beta_x-i\beta_y)^{-\nu}$ can be 
tested by not including some data points of the smallest system sizes. 
Although our dataset of five data points is not enough for a systematic 
analysis, we can still compare the one from the full dataset with 
the other from the reduced dataset excluding $L=8$. 
As shown in Table~\ref{tab:beta_c}, the locations of $\beta_c$ 
at the upper transitions are stable with the exclusion of $L=8$. 
At the lower transitions, the finite-size effect is much stronger 
as suggested by the test of $\psi_L$ in Fig.~\ref{fig6}, however 
the change of $\beta_c$ is still in the range of the values reported 
in the previous studies. In addition, the same procedures provides 
$\beta_c \approx 1.115$ in the $XY$ model for our leading-zero dataset 
of $L \le 128$, which is also comparable to the high-precision 
estimate $\beta_c \approx 1.1199$ given by 
the large-scale MC simulations~\cite{Hasenbusch2005,Komura2012}.

\begin{table}
\caption{\label{tab:beta_c}
Comparisons with the previous estimates of the transition
points for the upper ($\beta_c^\mathrm{high}$) and 
lower ($\beta_c^\mathrm{low}$) transitions in the $p$-state clock model. 
The last four rows indicate our estimates from the fits 
to Eq.~\eqref{eq:complex_beta} with the HOTRG dataset 
for $L \ge L_\mathrm{min}$ at $D_c=60$ and $70$.}
\begin{ruledtabular}
\begin{tabular}{llllr}
$\beta_c^{\mathrm{high}}$($p=5$) & $\beta_c^{\mathrm{low}}$($p=5$) &
$\beta_c^{\mathrm{high}}$($p=6$) & $\beta_c^{\mathrm{low}}$($p=6$) & 
reference \\ \hline 
& & 1.088(12) & 1.47(4) & \cite{Challa1986}\\ 
& & 1.1111 & 1.4706 & \cite{Yamagata1991}\\ 
& & 1.1101(7) & 1.4257(22) & \cite{Tomita2002}\\
1.0510(10) & 1.1049(10) & & & \cite{Borisenko2011}\\ 
& & 1.1086(6) & & \cite{Baek2013}\\ 
1.0593 & 1.1013 &  1.106(6) & 1.4286(82) & \cite{Kumano2013}\\
1.058(19) & 1.094(14) & & & \cite{Chatelain2014}\\
1.0504(1) & 1.1075(1) & & & \cite{Chen2018}\\
\hline
& & & & $D_c = 60$\\
1.060(2) & 1.097(3) & 1.110(6) & 1.436(6) & $L_\mathrm{min} = 8$\\
1.058(2) & 1.096(16) & 1.114(15) & 1.433(20) & $L_\mathrm{min} = 16$\\
\hline
& & & & $D_c = 70$\\
1.059(1) & 1.096(7) & 1.106(1) & 1.441(5) & $L_\mathrm{min} = 8$\\ 
1.058(1) & 1.101(6) & 1.106(2) & 1.444(2) & $L_\mathrm{min} = 16$\\ 
\end{tabular}
\end{ruledtabular}
\end{table}

\begin{figure}[t]
\centering\includegraphics[width=0.48\textwidth]{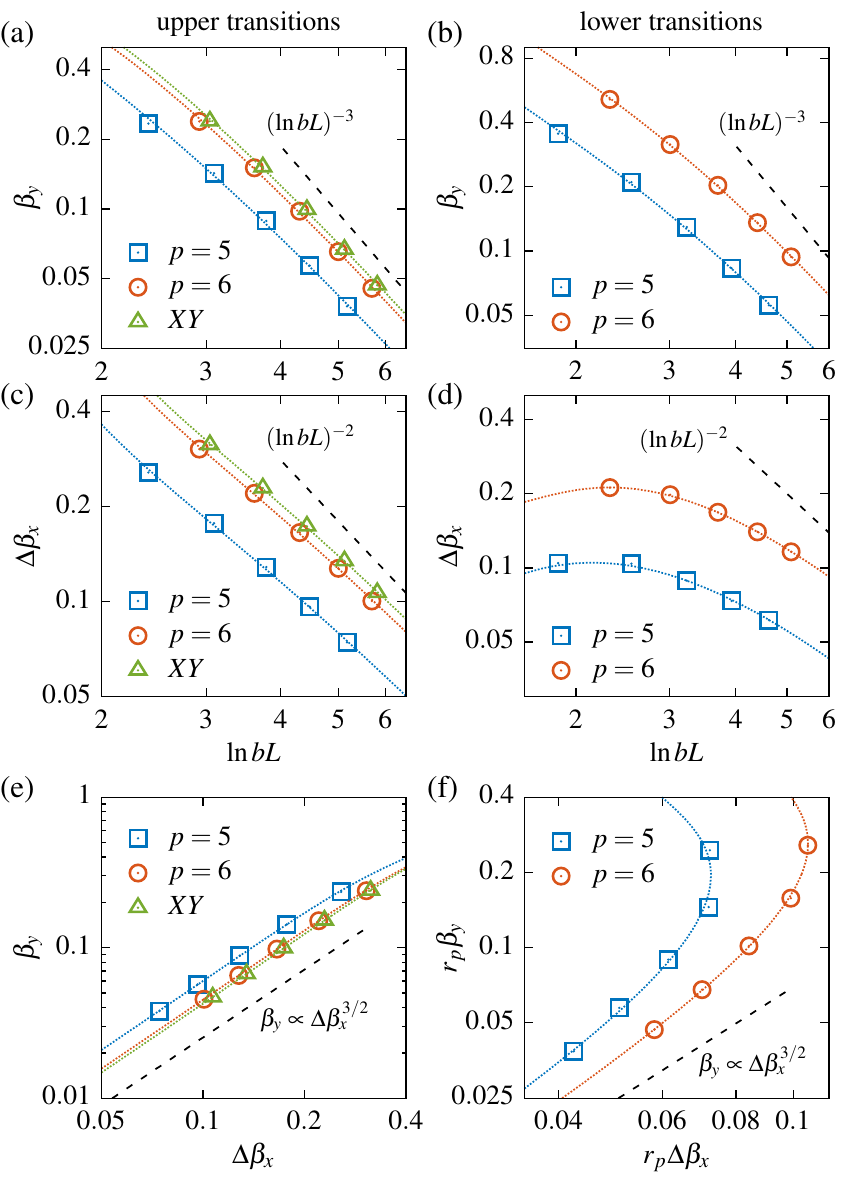}
\caption{\label{fig8}
Finite-size-scaling behaviors of the leading zeros at the transitions 
in the $p$-state clock model and the $XY$ model. 
The imaginary [(a) and (b)] and real [(c) and (d)] parts of the leading
zeros and their trajectory in the complex plane [(e) and (f)] 
are examined at the upper [(a), (c), and (e)] and lower [(b), (d) and (f)]
transitions. The dotted lines indicates 
Eqs.~\eqref{eq:fss_exact_beta_x} and \eqref{eq:fss_exact_beta_y}. 
The trajectories at the lower transitions are rescaled with 
the factor $r_p \equiv 1-\cos{(2\pi/p)}$ for visualization.
The data at $D_c=70$ and the parameters determined 
from Fig.~\ref{fig5} are used.
}
\end{figure}

The imaginary and real part of the leading zeros are expected 
to behave as $\beta_y \sim (\ln bL)^{-1-1/\nu}$ and 
$\Delta\beta_x \sim (\ln bL)^{-1/\nu}$ from the BKT ansatz
in the limit of a very large $L$.
While at a finite $L$, the leading zero behaviors deviate from 
these asymptotes, Eqs.~\eqref{eq:fss_exact_beta_x} and
\eqref{eq:fss_exact_beta_y} that include the finite-size corrections 
describe very well the FSS behaviors of the leading zeros, 
including an arclike trajectory at the lower transition,
as demonstrated in Fig.~\ref{fig8}.
The parameters in $\psi_L$ are determined from the linear fits 
shown in Fig.~\ref{fig5} in the procedures of locating $\beta_c$. 
The excellent agreement between the data points and the analytic
predictions again emphasizes that the logarithmic correction 
to the BKT ansatz is essential to the analysis of the leading zeros 
in the $p$-state clock model. 

Finally, let us discuss briefly the BKT exponent $\nu$ that is
fixed at the standard value of $\nu=1/2$ in our FSS tests. 
While it is hard to determine $\nu$ directly from 
$\Delta\beta_x$ or $\beta_y$ because of other fitting parameters 
being involved, the asymptote of the leading zero trajectory, 
$\beta_y \propto \Delta\beta_x^{1+1/\nu}$, can be checked for 
the consistency with the HOTRG data because it is free from 
the parameters other than the estimate of $\beta_c$. 
It turns out that at the upper transition, where the finite-size 
influence is less pronounced, the leading zero data agree very well 
with the prediction of $\beta_y \propto \Delta\beta_x^{3/2}$ 
as shown in Fig.~\ref{fig8}(e).
At the lower transitions, while the finite-size influence is much 
stronger as shown in Fig.~\ref{fig8}(f), the extrapolation of 
the leading-zero data points is getting closer to the asymptotic 
behavior of $\nu=1/2$ as $L$ increases, suggesting that 
the lower transitions as well as the upper ones are indeed of 
the same BKT type yet with the different appearance of 
the finite-size effects.

\section{Summary and conclusions}
\label{sec:summary}

We have investigated the numerical feasibility of the FSS analysis
with leading Fisher zeros for the BKT transitions and proposed 
the logarithmic corrections to the FSS ansatz to examine 
the two phase transitions in $p$-state clock models 
for $p=5$ and $6$ in the square lattices.
Our main findings can be summarized in two parts. 
(i) The reliability of the leading zero identification 
under finite MC noises highly depends on the type of the associated 
phase transition.
(ii) The combination of the HOTRG method and the logarithmic
correction to the FSS ansatz allows us to locate
the transition points, resolving the discrepancy between
the previous Fisher zero study and the other estimates from 
different measures.

We have found that the numerical visibility of the leading zero
exhibits the characteristic system-size scaling that depends on 
the type of phase transition.
The analytic prediction and the numerical tests in the Potts, Ising, 
and clock model suggest that the leading zero identification is the most 
robust against the finite MC noises in the first-order transition. 
In the second-order transition with a non-negative specific heat 
exponent, the tolerance to the finite noises decreases slowly
with increasing system size. In the BKT transition, 
the tolerance decay turns out to be exponential,
emphasizing the necessity of an highly accurate partition function
evaluation in the search for the leading zero.  

Employing the deterministic HOTRG method, we have identified 
the leading zeros for system sizes up to $L=128$ in the $p$-state 
clock model. Although, it turns out that the logarithmic correction  
is essential to the characterization of the leading zero behavior. 
The logarithmic correction works as a guide to locate 
the transition points, providing the Fisher-zero estimates 
that are in good agreement with the other estimates from different measures.
In addition, our formulation of the FSS ansatz indicates 
that an arclike trajectory of the leading Fisher zeros can occur
if the finite-size influence is strong as indeed observed 
at the lower transitions.
Our results significantly extend the previous Fisher zero
studies of the $p$-state clock model~\cite{Hwang2009,Kim2017} 
by describing both of the upper and lower transitions 
within the same BKT scaling ansatz of the leading zeros
with the finite-size correction.

\begin{acknowledgments}
This work was supported from the Basic Science Research Program 
through the National Research Foundation of Korea funded by 
the Ministry of Education (NRF-2017R1D1A1B03034669)
and the Ministry of Science and ICT (NRF-2019R1F1A1063211)
and also by a GIST Research Institute (GRI) grant funded 
by the GIST.
\end{acknowledgments}


\begin{thebibliography}{}

\bibitem{Cardy1988}
\textit{Finite-size Scaling}, 
edited by J. L. Cardy (North-Holland, Amsterdam, 1988).

\bibitem{Fisher1971}
M. E. Fisher, in \textit{Critical Phenomena, Vol. 51 of Proceedings of the ``Enrico Fermi'' International School of Physics}, edited by M. S. Green (Academic Press, New York, 1971).

\bibitem{Fisher1972}
M. E. Fisher and M. N. Barber, Phys. Rev. Lett. \textbf{28}, 1516 (1972).

\bibitem{Berche2012}
B. Berche, R. Kenna, and J.-C. Walter, Nucl. Phys. B \textbf{865}, 115 (2012).

\bibitem{Kenna2013a}
R. Kenna and B. Berche, Condens. Matter Phys. \textbf{16}, 23601 (2013). 

\bibitem{Kenna2015}
R. Kenna and B. Berche, in \textit{Order, Disorder and Criticality}, edited by Y. Holovatch (World Scientific, Singapore, 2015), Vol. 4, chap. 1.

\bibitem{Berezinskii1971} 
V. L. Berezinskii, Zh. Eksp. Teor. Fiz. \textbf{59}, 907 (1971); 
[Sov. Phys. JETP \textbf{32}, 493 (1971)].

\bibitem{Kosterlitz1972} 
J. M. Kosterlitz and D. Thouless, J. Phys. C \textbf{5}, L214 (1972).

\bibitem{Kosterlitz1973} 
J. M. Kosterlitz and D. Thouless, J. Phys. C \textbf{6}, 1181 (1973).

\bibitem{Bena2005}
I. Bena, M. Droz, and A. Lipowski, Int. J. Mod. Phys. B \textbf{19}, 4269 (2005).

\bibitem{Yang1952} 
C. N. Yang and T. D. Lee, Phys. Rev. \textbf{87}, 404 (1952).

\bibitem{Fisher1965} 
M. E. Fisher, in \textit{Lectures in Theoretical Physics},
edited by W. E. Brittin (University of Colorado Press, Boulder, 1965), 
Vol. 7C, chap. 1.

\bibitem{Janke2001} 
W. Janke and R. Kenna, J. Stat. Phys. \textbf{102}, 1211 (2001).

\bibitem{Kenna1995}
R. Kenna and A. C. Irving, Phys. Lett. B \textbf{351}, 273 (1995).

\bibitem{Irving1996}
A. C. Irving and R. Kenna, Phys. Rev. B \textbf{53}, 11568 (1996).

\bibitem{Kenna1997}
R. Kenna and A. C. Irving, Nucl. Phys. B \textbf{485}, 583 (1997).

\bibitem{Denbleyker2014}
A. Denbleyker, Y. Liu, Y. Meurice, M. P. Qin, T. Xiang, 
Z. Y. Xie, J. F. Yu, and H. Zou, 
Phys. Rev. D \textbf{89}, 016008 (2014);
H. Zou, Ph.D. thesis, University of Iowa (2014).

\bibitem{Rocha2016}
J. C. S. Rocha, L. A. S. M\'ol, and B. V. Costa, 
Comput. Phys. Commun. \textbf{209}, 88 (2016).

\bibitem{Costa2017}
B. V. Costa, L. A. S. M\'ol, and J. C. S. Rocha, 
Comput. Phys. Commun. \textbf{216}, 77 (2017).

\bibitem{Hwang2009} 
C.-O. Hwang, Phys. Rev. E \textbf{80}, 042103 (2009).

\bibitem{Kim2017}
D.-H. Kim, Phys. Rev. E \textbf{96}, 052130 (2017). 

\bibitem{Wang2001a}
F. Wang and D. P. Landau, Phys. Rev. Lett. \textbf{86}, 2050 (2001).

\bibitem{Wang2001b}
F. Wang and D. P. Landau, Phys. Rev. E \textbf{64}, 056101 (2001).

\bibitem{Alves2002}
N. A. Alves, J. P. N. Ferrite, and U. H. E. Hansmann,
Phys. Rev. E \textbf{65}, 036110 (2002).

\bibitem{Kenna2006}
R. Kenna, Condens. Matter Phys. \textbf{9}, 283 (2006).

\bibitem{Hasenbusch2005}
M. Hasenbusch, J. Phys. A: Math. Gen. \textbf{38}, 5869 (2005).

\bibitem{Komura2012}
Y. Komura and Y. Okabe, J. Phys. Soc. Jpn. \textbf{81}, 13001 (2012).

\bibitem{Kenna2006a}
R. Kenna, D. A. Johnston, and W. Janke, Phys. Rev. Lett \textbf{96}, 115701 (2006).

\bibitem{Kenna2006b}
R. Kenna, D. A. Johnston, and W. Janke, Phys. Rev. Lett. \textbf{97}, 155702 (2006).

\bibitem{Kenna2013}
R. Kenna, in \textit{Order, Disorder and Criticality}, edited by Y. Holovatch (World Scientific, Singapore, 2013), Vol. 3, chap. 1.

\bibitem{book:Jose}
{\it 40 Years of Berezinskii-Kosterlitz-Thouless Theory}, 
edited by J. V. Jos\'e (World Scientific, London, 2013).

\bibitem{Elitzur1979}
S. Elitzur, R. B. Pearson, and J. Shigemitsu, 
Phys. Rev. D \textbf{19}, 3698 (1979).

\bibitem{Cardy1980}
J. L. Cardy, J. Phys. A: Math. Gen. \textbf{13}, 1507 (1980).

\bibitem{Einhorn1980}
M. B. Einhorn, R. Savit, and E. Rabinovici, 
Nucl. Phys. B \textbf{170}, 16 (1980).

\bibitem{Hamer1980}
C. J. Hamer and J. B. Kogut, Phys. Rev. B \textbf{22}, 3378 (1980).

\bibitem{Frohlich1981}
J. Fr\"ohlich and T. Spencer, Comm. Math. Phys. \textbf{81}, 527 (1981).

\bibitem{Nienhuis1984}
B. Nienhuis, J. Stat. Phys. \textbf{34}, 731 (1984).

\bibitem{Ortiz2012}
G. Ortiz, E. Cobanera, and Z. Nussinov, Nucl. Phys. B \textbf{854}, 780 (2012).

\bibitem{Tobochnik1982}
J. Tobochnik, Phys. Rev. B \textbf{26}, 6201 (1982); \textbf{27}, 6972 (1983).

\bibitem{Challa1986}
M. S. S. Challa and D. P. Landau, Phys. Rev. B \textbf{33}, 437 (1986).

\bibitem{Yamagata1991} A. Yamagata and I. Ono, 
J. Phys. A: Math. Gen. \textbf{24}, 265 (1991).

\bibitem{Tomita2002}
Y. Tomita and Y. Okabe, Phys. Rev. B {\bf 65}, 184405 (2002).

\bibitem{Borisenko2011}
O. Borisenko, G. Cortese, R. Fiore, M. Gravina, and A. Papa, 
Phys. Rev. E {\bf 83}, 041120 (2011).

\bibitem{Borisenko2012}
O. Borisenko, V. Chelnokov, G. Cortese, R. Fiore, M. Gravina, and A. Papa,
Phys. Rev. E {\bf 85}, 021114 (2012).

\bibitem{Lapilli2006}
C. M. Lapilli, P. Pfeifer, and C. Wexler, Phys. Rev. Lett. {\bf 96}, 140603 (2006).

\bibitem{Baek2010a}
S. K. Baek, P. Minnhagen, and B. J. Kim, Phys. Rev. E {\bf 81}, 063101 (2010).

\bibitem{Baek2010b}
S. K. Baek, P. Minnhagen, Phys. Rev. E {\bf 82}, 031102 (2010).

\bibitem{Baek2013}
S. K. Baek, H. M\"akel\"a, P. Minnhagen, and B. J. Kim, 
Phys. Rev. E {\bf 88}, 012125 (2013).

\bibitem{Kumano2013}
Y. Kumano, K. Hukushima, Y. Tomita, and M. Oshikawa, 
Phys. Rev. B {\bf 88}, 104427 (2013).

\bibitem{Chatelain2014}
C. Chatelain, J. Stat. Mech. (2014) P11022.

\bibitem{Chen2017}
J. Chen, H.-J. Liao, H.-D. Xie, X.-J. Han, R.-Z. Huang, S. Cheng,
Z.-C. Wei, Z.-Y. Xie, and T. Xiang,
Chin. Phys. Lett. \textbf{34}, 050503 (2017).

\bibitem{Chen2018}
Y. Chen, Z.-Y. Xie, and J.-F. Yu, Chin. Phys. B \textbf{27}, 080503 (2018).

\bibitem{Surungan2019}
T. Surungan, S. Masuda, Y. Komura, and Y. Okabe, 
J. Phys. A: Math. Theor. \textbf{52}, 275002 (2019).

\bibitem{Falcioni1982} 
M. Falcioni, E. Marinari, M. L. Paciello, G. Parisi, and B. Taglienti, 
Phys. Lett. B \textbf{108}, 311 (1982).

\bibitem{Marinari1984}
E. Marinari, Nucl. Phys. B \textbf{235}, 123 (1984).

\bibitem{Ferrenberg1988}
A. M. Ferrenberg and R. H. Swendsen, 
Phys. Rev. Lett. \textbf{61}, 2635 (1988).

\bibitem{Ferrenberg1989}
A. M. Ferrenberg and R. H. Swendsen,
Phys. Rev. Lett. \textbf{63}, 1195 (1989).

\bibitem{Alves1992} 
N. A. Alves, B. A. Berg, and S. Sanielevici, 
Nucl. Phys. B \textbf{376}, 218 (1992).

\bibitem{mpsolve}
D. A. Bini and L. Robol, 
J. Comput. Appl. Math. \textbf{272}, 276 (2014).

\bibitem{Vogel2013}
T. Vogel, Y. W. Li, T. W\"ust, and D. P. Landau,
Phys. Rev. Lett. \textbf{110}, 210603 (2013).

\bibitem{Vogel2014}
T. Vogel, Y. W. Li, T. W\"ust, and D. P. Landau,
Phys. Rev. E \textbf{90}, 023302, (2014).

\bibitem{Valentim2015}
A. Valentim, J. C. S. Rocha, S.-H. Tsai, Y. W. Li, 
M. Eisenbach, C. E. Fiore, and D. P. Landau,
J. Phys.: Conf. Ser. \textbf{640}, 012006 (2015).

\bibitem{Ren2016}
Y. Ren, S. Eubank, and M. Nath, Phys. Rev. E \textbf{94}, 042125 (2016).

\bibitem{Chan2017a}
C. H. Chan, G. Brown, and P. A. Rikvold, Phys. Rev. E \textbf{95}, 053302 (2017).

\bibitem{Chan2017b}
C. H. Chan, G. Brown, and P. A. Rikvold, Phys. Rev. B \textbf{96}, 174428 (2017).

\bibitem{Xie2012}
Z. Y. Xie, J. Chen, M. P. Qin, J. W. Zhu, L. P. Yang, and T. Xiang, 
Phys. Rev. B \textbf{86}, 045139 (2012).

\bibitem{Levin2007}
M. Levin and C. P. Nave, Phys. Rev. Lett. \textbf{99}, 120601 (2007).

\bibitem{Xie2009}
Z. Y. Xie, H. C. Jiang, Q. N. Chen, Z. Y. Weng, and T. Xiang, 
Phys. Rev. Lett. \textbf{103}, 160601 (2009).

\bibitem{Zhao2010}
H. H. Zhao, Z. Y. Xie, Q. N. Chen, Z. C. Wei, J. W. Cai, and T. Xiang,
Phys. Rev. B \textbf{81}, 174411 (2010).

\bibitem{GarciaSaez2015}
A. Garc\'ia-Saez and T.-C. Wei, Phys. Rev. B \textbf{92}, 125132 (2015).

\bibitem{Wang2014}
S. Wang, Z.-Y. Xie, J. Chen, B. Normand, and T. Xiang, 
Chin. Phys. Lett. \textbf{31}, 070503 (2014).

\bibitem{Yu2014}
J. F. Yu, Z. Y. Xie, Y. Meurice, Y. Liu, A. Denbleyker, H. Zou, 
M. P. Qin, J. Chen, and T. Xiang, 
Phys. Rev. E \textbf{89} 013308 (2014).

\bibitem{supp}
See Supplemental Material for the table of the numerical data
of the leading zeros identified.

\bibitem{Denbleyker2007}
A. Denbleyker, D. Du, Y. Meurice, and A. Velytsky, 
Phys. Rev. D \textbf{76}, 116002 (2007).

\bibitem{denNijs1979} 
M. P. M. den Nijs, J. Phys. A, \textbf{12}, 1857 (1979).

\bibitem{Nagai2013}
T. Nagai, Y. Okamoto, and W. Janke,
Condens. Matter Phys. \textbf{16}, 23605 (2013).

\end{thebibliography}
\end{document}